\documentclass[prl,twocolumn,nofootinbib]{revtex4}
\usepackage{amsmath,epsfig,bm}
\usepackage{subfigure}

\newcommand\pd{\partial}

\newcommand\paper{paper}
\newcommand\p{{\bm{p}}}
\newcommand\la{\langle}
\newcommand\ra{\rangle}
\newcommand\xis{\xi}
\newcommand\cO{{\cal O}}

\begin{document}

\title{On the sign of kurtosis near the QCD critical point}

\author{M.~A.~Stephanov}
\affiliation{Department of Physics, University of Illinois, Chicago, 
Illinois 60607, USA}

\begin{abstract}
  We point out that the quartic cumulant (and kurtosis) of the
  order parameter fluctuations is universally {\em negative} when the critical
  point is approached on the crossover side of the phase separation
  line. As a consequence, the kurtosis of a
  fluctuating observable, such as, e.g., proton multiplicity, may
  become smaller than the value given by independent Poisson
  statistics. We discuss implications for the Beam Energy Scan program
  at RHIC.
\end{abstract}

\maketitle

\section{Introduction}
\label{sec:intro}

Mapping the QCD phase diagram as a function of temperature $T$ and
baryochemical potential $\mu_B$ is one of the fundamental goals of
heavy-ion collision experiments. QCD critical point is a distinct
singular feature of the phase diagram. It is a ubiquitous property of
QCD models based on the chiral symmetry breaking
dynamics %~\cite{Asakawa:1989bq,Barducci:1989wi} 
(see, e.g.,
 Ref.\cite{Stephanov:2004wx} for a review and further 
references). Locating
the point using first-principle lattice calculations is a formidable
challenge %~\cite{Fodor:2001pe,Ejiri:2003dc,Gavai:2004sd,Gavai:2008zr,deForcrand:2003bz}.
%Recent progress and results are encouraging, but much work needs to be
%done to understand and constrain systematic errors (see,
%e.g., %Refs.\cite{Schmidt:2009qq,Gupta:2009mu,Philipsen:2009dn} and
%reviews~\cite{%Schmidt:2006us,Stephanov:2007fk,
(see, e.g., Ref.\cite{deForcrand:2010ys} 
for a recent review and references).  
If the critical point is situated in the
region accessible to heavy-ion collision experiments it can be
discovered experimentally. The search for the critical point is
planned at the Relativistic Heavy Ion Collider (RHIC) at BNL, the
Super Proton Synchrotron (SPS) at CERN, the future Facility for
Antiproton and Ion Research (FAIR) at GSI, and Nuclotron-based Ion
Collider Facility (NICA) in
Dubna (see, e.g., Ref.\cite{Mohanty:2009vb}).

The characteristic feature of a critical point is the divergence of
the correlation length $\xi$ and of the magnitude of the fluctuations.
The simplest measures of fluctuations in heavy-ion collisions are the
variances of the event-by-event observables such as multiplicities or
mean transverse momenta of particles. The singular, critical
contribution to these variances diverges as (approximately) $\xi^2$,
and would manifest in a non-monotonic dependence
of such measures as the critical point is passed by during the beam
energy scan~\cite{Stephanov:1998dy,Stephanov:1999zu}. In realistic
heavy ion collision the divergence of $\xi$ is cut-off by the effects
of critical slowing down \cite{Stephanov:1999zu,Berdnikov:1999ph}, and
the estimates of the maximum correlation length are in the range of at
most $2-3$ fm, compared to the natural $0.5-1$ fm away from the
critical point.  However, higher, non-Gaussian, moments of the
fluctuations depend much more sensitively on $\xi$, according
to~Ref.\cite{Stephanov:2008qz}. For example, the 4-th moment grows as
$\xi^7$ near the critical point, making it an attractive experimental
tool.  In this \paper\ we follow up on the results of
Ref.\cite{Stephanov:2008qz} to point out that the sign of the 4-th
moment could be negative as the critical point is approached from the
crossover side of the QCD phase transition.

The sign of various moments have been discussed in the literature in
related contexts: see, e.g., discussion of the sign of
the 3-rd moment in~Ref.\cite{Asakawa:2009aj} or the 6-th and 8-th
moments in~Ref.\cite{Friman:2011pf} and also numerical lattice
calculations in Ref.\cite{Gavai:2010zn} where the possible sign change
of kurtosis is noted.  

In this \paper\ we shall address specifically the sign of the 4-th
moment (or kurtosis) and do it in a more universal and quantitative
way than has been done previously, by using the known parametric form
of the universal equation of state near the critical point. We
emphasize universality of the behavior of the kurtosis and draw
experimental consequences from these results.

\section{Kurtosis and universal effective potential}
\label{sec:kurt-univ-effect}

Let us begin, as in Ref.\cite{Stephanov:2004wx}, by describing
fluctuations of the order parameter field $\sigma(\bm x)$ near a
critical point using the probability distribution
\begin{equation}
  \label{eq:P-Omega-sigma}
  P[\sigma] \sim \exp\left\{ - \Omega[\sigma]/T\right\},
\end{equation}
where $\Omega$ is the effective action (free energy) functional for
the field $\sigma$, which can be expanded in powers of $\sigma$ as
well as in the gradients (we chose $\sigma=0$ at the minimum):
\begin{equation}
  \label{eq:Omega-sigma}
  \Omega%[\sigma] 
=\! \int\!d^3\bm x\left[
\frac{(\bm\nabla\sigma)^2}{2} +
\frac{m_\sigma^2}2 \sigma^2 
+ \frac{\lambda_3}{3}\sigma^3
+ \frac{\lambda_4}{4}\sigma^4 + \ldots
\right]
\,.
\end{equation}
Calculating 2-point correlator $\la\sigma(\bm x)\sigma(0)\ra$ we
find that the correlation length $\xis=m_\sigma^{-1}$.
For the moments of the zero momentum mode $\sigma_V\equiv
\int\!d^3x\,\sigma(x)$
in a system of volume $V$ we find at tree level
\begin{equation}
  \label{eq:sigma-moments}
  \begin{split}
  &  \kappa_2=\la \sigma_V^2 \ra = V T\,\xis^2\,;
%\\
%  &
\qquad
\kappa_3=\la \sigma_V^3 \ra = {2 \lambda_3 VT^2}\, \xis^6\,;
\\
  &\kappa_4=\la \sigma_V^4 \ra_c
= {6VT^3}\, [\,2(\lambda_3\xis)^{2} - \lambda_4\,]\, \xis^8\,.   
  \end{split}
\end{equation} 
where $\la\sigma_V^4 \ra_c \equiv \la \sigma_V^4 \ra - 3\la \sigma_V^2
\ra^2$ denotes the connected 4-th central moment (the
4-th cumulant). The critical point is characterized by $\xis\to\infty$. The
central observation in Ref.\cite{Stephanov:2008qz} was that the
higher moments (cumulants) $\kappa_3$ and $\kappa_4$ diverge with
$\xis$ much faster than the quadratic moment $\kappa_2$. Here we shall
 point out that the {\em sign} of the 4-th moment
$\kappa_4$ is negative in a certain sector near the critical
point. More precisely, the 4-th cumulant is negative when the
critical point is approached from the crossover side.
Let us demonstrate this in several complementary ways.

A simple way to see why the kurtosis is negative is by following the
evolution of the probability distribution of $\sigma_V$ as we approach
the critical point along the crossover line. In Ising scaling
coordinates: along $H=0$, $t>0$ ray. Away from the critical point,
more precisely for $\xi^3\ll V$, the central limit theorem dictates
that the probability distribution of $\sigma_V$ is Gaussian, with a
vanishingly small kurtosis. As we approach the critical point
the distribution develops non-Gaussian shape. This intermediate shape
is a deformation of the Gaussian towards a two-peak distribution,
corresponding to the phase coexistence on the opposite, first-order
transition side ($t<0$) of the critical point. Such a shape is clearly
less ``peaked'' than the Gaussian, and thus corresponds to negative
kurtosis.

More quantitatively, the kurtosis vanishes as $1/V$ at (almost) any
point away from the critical point, i.e.,
\begin{equation}
  \label{eq:K-def}
  K\equiv
\kappa_4/\kappa_2^2=\cO(\xi^3/V).
\end{equation}
The exception is the coexistence line ($H=0$, $t<0$ ray). The
distribution there has two peaks of equal height and its kurtosis is
$K=-2+\cO(\xi^3/V)$.\footnote{ In fact, this transition of the shape
  of the distribution around a critical point is universal to all
  critical points of the same (Ising model) universality class. At
  finite $V$, at $t=0$, the value of $K$ is independent of $V$ (the
  correlation length $\xi\sim V^{1/3}$ is as large as it can be at
  given volume). This value of $K$ is a universal number (it depends
  only on the boundary conditions for a given universality class) and
  is well-known. It is usually expressed as the value of the Binder
  cumulant $B_4\approx 1.6$, which means $K=B_4-3\approx -1.4<0$.} 
It is important to note that this is only true strictly {\em on\/} the
coexistence line $H=0$, for the moments
measured around the symmetric point of the probability distribution of
$\sigma_V$, which is actually a dip, not a peak, for $t<0$. At any
point close to the coexistence line, i.e., at $H\neq 0$, $t<0$, the
kurtosis around the dominant peak is positive.

In the scaling regime  (close to, but not 
at the critical point) where  $\xi$ is much greater than the
microscopic scale, $a$, but still much less than the linear size of the system:
$a\ll \xi\ll V^{1/3}$, the coefficient of $\xi^3/V$ in
Eq.~(\ref{eq:K-def}) can be expressed in terms of the
couplings $\lambda_i$ using Eqs.~(\ref{eq:sigma-moments}): 
\begin{equation}
  \label{eq:K-lambda}
  K = 6\left(\,2\lambda_3^2\xi^3-\lambda_4\xi\,\right)\,\frac{\xi^3}{V}\,.
\end{equation}
These
couplings, and in fact the shape of the effective potential, is also
universal. In particular, $\lambda_4$
scales with~$\xi$ as 
$\lambda_4=\tilde\lambda_4\xi^{-1}$, where the universal value of
$\tilde\lambda_4$ is known approximately to be 4.0 on the crossover
line (see, e.g.,
Ref.\cite{Tsypin} for a review).\footnote{As in Ref.\cite{Stephanov:2008qz}, for
  simplicity and consistency with our overall level of precision, we neglect the anomalous scaling dimension $\eta$, which
  is only of order few percent.} Since on the
crossover line $\lambda_3=0$ and $\lambda_4>0$, 
it is clear from Eq.~(\ref{eq:sigma-moments}) that $K<0$. 

Away from the crossover line ($H=0$, $t>0$ ray) the distribution is skewed:
$\lambda_3\neq0$. This makes the kurtosis positive, according to
Eq.~(\ref{eq:K-lambda}), except for a
certain sector around the crossover line.

\section{The universal equation of state around the critical point}

\label{sec:univ-equat-state}

To extend this analysis away from the crossover line, i.e., to take into account
$\lambda_3\neq0$ in Eq.~(\ref{eq:K-lambda}), we need to know the equation of state, in
particular, $\kappa_4$ as a function of both Ising variables: reduced
temperature $t$ and magnetic field $H$. In the scaling regime near
$t=H=0$ this equation of state is also universal. For the Ising model
it is known to order $\varepsilon^3$ in the epsilon expansion as well
as numerically.

Before we discuss this universal form, let us keep in mind that the
mapping of QCD phase diagram in the $T$, $\mu_B$ plane into
$t$, $H$ plane is not universal. However, this mapping
is analytic, i.e., both functions $t(T,\mu_B)$ and $H(T,\mu_B)$ are
analytic at the critical point, which is mapped into the origin,
$t(T^{\rm cp},\mu_B^{\rm cp})=H(T^{\rm cp},\mu_B^{\rm cp})=0$.

The standard parametrization, Ref.\cite{Zinn-Justin}, of the equation
of state in the scaling domain near the critical point is in terms of
two new scaling variables $R$ and $\theta$ (it has been applied in
the context of QCD before, Ref.\cite{Nonaka:2004pg}). Denoting
the ``magnetization'' by $M=\la\sigma_V\ra/V$ we define $R$ and
$\theta$ as
\begin{align}
  \label{eq:param}
M = R^\beta\theta,
\quad
t = R (1-\theta^2),
\end{align}
Then the equation of state can be expressed in terms of the single
function $h(\theta)$ as
\begin{equation}
  \label{eq:H-h}
  H = R^{\beta\delta}h(\theta)\,.
\end{equation}
Unlike the explicit form of the singular equation of state $M=M(t,H)$,
the function $h(\theta)$ is analytic. It has two zeros. One, at
$\theta=0$, corresponds to the crossover line ($t>0$, $H=0$ ray), another,
at some $\theta=\theta_1>1$, corresponds to the coexistence
(first-order transition) line ($t<0$, $H=0$ ray).  The function $h(\theta)$ must
also be odd since $M(-H)=-M(H)$.  The simplest function obeying all
these requirements is a cubic polynomial
\begin{equation}
  \label{eq:h-cubic}
  h(\theta) = \theta(2-3\theta^2).
\end{equation}
where the value $\theta_1=\sqrt{3/2}$ is a good approximation to the
universal value for the Ising model (and
correct up to $\cO(\varepsilon^2)$).
The choice (\ref{eq:h-cubic}) is known as  the
linear parametric model, Ref.\cite{Schofield:1969}. It
describes the equation of state with precision quite sufficient for 
our purposes. The linear parametric model
is also known to be exact up to $\cO(\varepsilon^3)$.%, Ref.\cite{Zinn-Justin}. 

\begin{figure}[h]
%  \centering
\subfigure[]{
\includegraphics[height=20em]{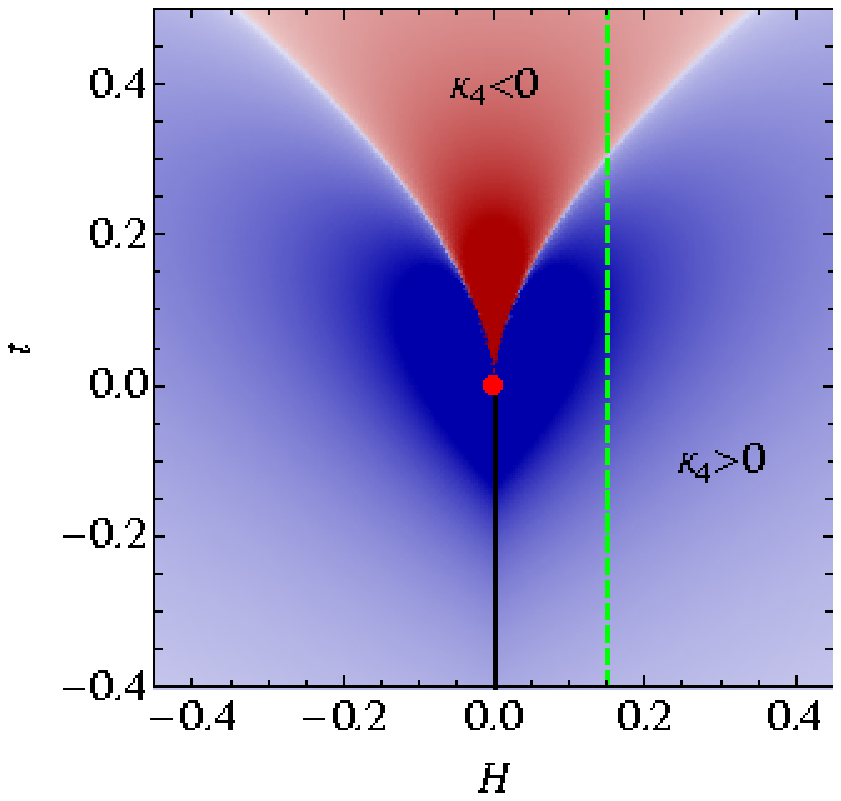}
\label{fig:k4-dp}
}
%\begin{minipage}[b]{.45\linewidth}
%\subfigure[]{
%\includegraphics[height=14em]{pdqcd-fo-2.eps}
%\label{fig:qcd-pd}
%}
\subfigure[]{
\includegraphics[height=11em]{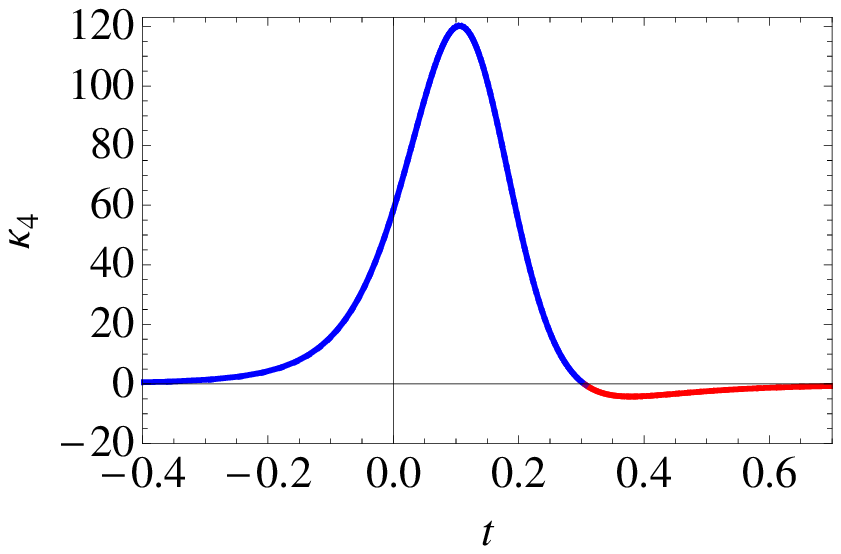}
\label{fig:k4-t}
}
%\end{minipage}
\caption[]{%Left:
(color online)
  (a) -- the density plot of the function $\kappa_4(t,H)$ given by
  Eq.~(\ref{eq:kappa4-theta-R}) obtained using Eq.~(\ref{eq:k4-MH})
  for the linear parametric model
  Eqs.~(\ref{eq:param}),~(\ref{eq:H-h}),~(\ref{eq:h-cubic}) and
  $\beta=1/3$, $\delta=5$. The $\kappa_4<0$ region is red, the
  $\kappa_4>0$ -- is blue.  (b) -- the dependence of
  $\kappa_4$ on $t$ along the vertical dashed green line on the density plot
  above. This line is the simplest example of a possible mapping of
  the freezeout curve (see Fig.~\ref{fig:qcd-pd}).  The units of $t$,
  $H$ and $\kappa_4$ are arbitrary.}
\label{fig:k4-th}
\end{figure}

Using this parametric equation of state,
we can calculate the moments by taking derivatives at fixed $t$,
%, starting with
% \begin{multline}
%   \label{eq:k2}
%   \kappa_2\left(t(R,\theta),\,H(R,\theta)\right) 
% = \left(\frac{\pd M}{\pd H}\right)_t
% \\
% =\frac{d M\left(R=t/(1-\theta^2),\,\theta\right)/d\theta}
% {d H\left(R=t/(1-\theta^2),\,\theta\right)/d\theta},
% \end{multline}
up to an overall normalization, unimportant in the present context (it
can be fixed by Eq.~(\ref{eq:sigma-moments})).
%Repeating this partial differentiation twice more we obtain
In particular,
\begin{equation}
  \label{eq:k4-MH}
  \kappa_4(t,H) = \left(\frac{\pd^3 M}{\pd H^3}\right)_t\,.
\end{equation}
%The result is cumbersome, but is easily handled by a symbolic
%manipulation program. 
For our purposes, it would be sufficient to use
the approximate rational values of critical exponents $\beta=1/3$ and
$\delta=5$, which are within few percent of their exact values in
three dimensions. The result of Eq.~(\ref{eq:k4-MH}) can then be
simplified to
\begin{equation}
\kappa_4(t,H)=
- 12\, \frac{81 - 783 \theta^2 + 105 \theta^4 - 5 \theta^6 + 
   2 \theta^8}{R^{14/3} (3 - \theta^2)^3 (3 + 2 \theta^2)^5}\,.
\label{eq:kappa4-theta-R}   
\end{equation}
We represent $\kappa_4(t,H)$  graphically as a
density plot in Fig.~\ref{fig:k4-th}. 
We see that the 4-th cumulant (and kurtosis) is negative
in the sector bounded by two curved rays
$H/t^{\beta\delta}=\pm{\rm const}$ (corresponding to $\theta\approx
\pm 0.32$).

\begin{figure}
  \includegraphics[height=14em]{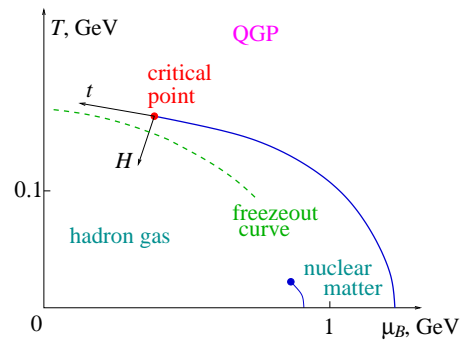}
\caption[]{A sketch of the phase diagram of QCD
  with the freezeout curve and a possible mapping of the Ising
  coordinates $t$ and $H$.}
  \label{fig:qcd-pd}
\end{figure}

Also in Fig.~\ref{fig:k4-th} we show the dependence of $\kappa_4$
along a line
which could be thought of as representing a possible mapping of the
freezeout trajectory (Fig.~\ref{fig:qcd-pd}) onto the $tH$ plane. 
Although the absolute value
of the peak in $\kappa_4$ depends on the proximity of the
freezeout curve to the critical point, the ratio of the maximum to
minimum along such an $H={\rm const}$ curve is a universal number,
approximately equal to $-28$ from Eq.~(\ref{eq:kappa4-theta-R}).

The negative minimum is small relative to the positive peak, but given
the large size of the latter, Ref.\cite{Stephanov:2008qz,Athanasiou:2010kw}, the
negative contribution to kurtosis may be significant. In addition, the
mapping of the freezeout curve certainly need not be $H={\rm const}$,
and the relative size of the positive and negative peaks depends
sensitively on that.

The trend described above appears to show in the recent lattice data,
Ref.\cite{Gavai:2010zn}, obtained using Pade resummation of the
truncated Taylor expansion in $\mu_B$. As the chemical potential is
increased along the freezeout curve, the 4-th moment of the baryon
number fluctuations begins to decrease, possibly turning negative, as
the critical point is approached (see Fig.2 in
Ref.\cite{Gavai:2010zn}).

Another observation, which we shall return to at the end of the next section,
is that $-\kappa_4$ grows as we approach the crossover line,
corresponding to $H=0$, $t>0$ on the diagram in Fig.~\ref{fig:k4-dp}.
On the QCD phase diagram the freezeout point will move in this
direction if one reduces the size of the colliding nuclei or selects
more peripheral collisions (the freezeout occurs earlier, i.e., at
higher $T$, in a smaller system).

\section{Experimental observables}
\label{sec:exper-observ}

In this section we wish to connect the results for the fluctuations of
the order parameter field $\sigma$ to the fluctuations of the
observable quantities.  As an example we consider the fluctuations of
the multiplicity of given charged particles, such as pions or protons.

For completeness we shall briefly rederive the results of
Ref.\cite{Stephanov:2008qz} using a simple model of fluctuations. The model
captures the most singular term in the contribution of
the critical point to the fluctuation observables. Consider a given
species of particle interacting with fluctuating critical mode field
$\sigma$. The infinitesimal change of the field $\delta\sigma$ leads
to a change of the effective mass of the particle by the amount
$\delta m = g \delta \sigma$. This could be considered a definition of
the coupling $g$. For example, the coupling of protons in the sigma
model is $g\sigma \bar p p$.
%, and the coupling of pions (in
%the notation of Ref.\cite{Stephanov:1999zu}) $2 G\sigma\pi^+\pi^-$ gives
%$g_\pi=G/m_\pi$. 
The fluctuations $\delta f_\p$ of the momentum space
distribution function $f_\p$ consist of the pure statistical
fluctuations $\delta f_\p^0$ around the equilibrium distribution
$n_\p$ for a particle of a given mass, which itself fluctuates. This
gives
\begin{equation}
  \label{eq:3}
  \delta f_\p = \delta f_\p^0 + \frac{\pd n_\p}{\pd m}\, g\,\delta\sigma\,.
\end{equation}
Using this equation we can calculate the most singular contribution
from the critical fluctuations to the moments or correlators of
$\delta f_\p$. The fluctuation of the multiplicity $N=V\,d\int_\p f_\p$ is given by
\begin{equation}
  \label{eq:delta-N}
  \delta N = \delta N^0 + V\, g\, \delta\sigma\,d
\int_{\bm p}  \frac{\pd n_\p}{\pd m}\,,
\end{equation}
where $d$ is the degeneracy factor (e.g., number of spin or charge states of
the particle). Neglecting, for clarity and simplicity, the effects of quantum
statistics, i.e., assuming $n_\p\ll 1$, we can use  Poisson 
statistics for $\delta N^0$. Using additivity of the cumulants 
(their defining property), and assuming $\delta N^0$ and
$\delta\sigma$ are uncorrelated, the contribution of the critical
fluctuations can be expressed in terms of the corresponding moments of the
critical field $\sigma$ fluctuations. For example, 
the contribution to the 4-th moment can be expressed as 
(cf. Refs.\cite{Stephanov:2008qz,Athanasiou:2010kw})
\begin{equation}
  \label{eq:N4}
  \la(\delta N)^4\ra_c = \la N\ra + \la\sigma_V^4\ra_c
\left(\frac{g\,d}{T} \int_{\bm p}\frac{ n_{\bm p}
%(1\pm n_{\p})
}{\gamma_{\bm p}}\right)^4 + \ldots,
\end{equation}
where $\gamma_\p = (dE_\p/dm)^{-1}$ is the relativistic gamma-factor of a
particle with momentum $\p$ and mass~$m$.  The first term on the
r.h.s. of Eq.~(\ref{eq:N4}) is the Poisson contribution. We neglected
$n_\p\ll1$ in the quantum statistics factor $(1\pm n_\p)$ for
simplicity, and we denoted by ``$\ldots$'' other contributions, less
singular at the critical point. The model is admittedly crude, but it
illustrates the mechanism and correctly captures 
the most singular contribution near the critical point.

In the region near
the critical point where $\kappa_4=\la\sigma_V^4\ra_c$ is negative, the
4-th cumulant of the fluctuations will be smaller than its Poisson value,
$\la N\ra$. The measure defined in Ref.\cite{Stephanov:2008qz} as
$\omega_4(N)=\la (\delta N)^4\ra_c/\la N \ra$ will be less than 1. By
how much will depend sensitively on the correlation length (as
$\xi^7$), i.e., on how close the freezeout occurs to the critical
point, as well as on other factors (for protons, most significantly, 
on the value of $\mu_B$.)
We shall not attempt to estimate this effect quantitatively in this
\paper. The analysis of Ref.\cite{Athanasiou:2010kw} suggests, 
however, that this effect
for protons can be significant compared to the Poisson value 
already for $\xi\sim 2$ fm.

Usual caveats apply: other (non-trivial) contributions to
moments which do not behave singularly
at the critical point can turn out to be relatively large. It is
beyond the scope of the \paper\ to estimate these effects. The size of
these background contributions could, in principle, be determined
experimentally by performing measurements away from the critical point.

We conclude by asking an obvious question: has the effect of the
negative kurtosis been observed? Data from STAR indicate that at
$\sqrt s=19.6$ GeV the ratio $\kappa_4/\kappa_2$ might be
substantially smaller than its Poisson value 1, 
see Fig.~6 in Ref.\cite{Kumar:2011de},
while it is very close to 1 at higher $\sqrt s$
(smaller~$\mu_B$). Unfortunately, the statistics accumulated in the
short run at $\sqrt s=19.6$ GeV is clearly not sufficient to make a
reliable conclusion.  It would be interesting to see if this effect
persists with more statistics at this energy. If confirmed, this
result could indicate that the critical point is close, at somewhat
larger values of $\mu_B$ (smaller $\sqrt s$). In this case, as we
already discussed at the end of the previous section, the universality
would also predict that the negative kurtosis effect should increase
in more peripheral collisions at the same $\sqrt s$. At smaller values of
$\sqrt s$ the effect should change sign, increasing kurtosis above
its Poisson value.

\section*{Acknowledgments}

The author thanks O.~Evdokimov for discussions.
 This work is supported by the DOE grant No.\
DE-FG0201ER41195.

\end{document}